\documentclass[a4paper,11pt]{elsarticle}
\usepackage[colorlinks,bookmarksopen,bookmarksnumbered,citecolor=red,urlcolor=blue]{hyperref}

\usepackage{multibib}
\usepackage[T1]{fontenc}         
\usepackage[latin1]{inputenc}
\usepackage{graphicx}
\usepackage{amsmath,amssymb}
\usepackage{bm}
\usepackage{tikz}
\usepackage{color}
\usepackage{helvet}
\usepackage{courier}
\usepackage{makeidx}
\usepackage{footmisc}
\usepackage{type1cm}         
\usepackage{textcomp}
\usepackage{subfig}
\usepackage{enumitem,xcolor}    
\usepackage{yhmath}
\usepackage{esdiff}
\usepackage{mathtools}

\DeclareCaptionLabelSeparator{colon}{. }
\usepackage[labelfont={bf},labelsep={colon}]{caption}

\makeatletter
\DeclareRobustCommand\dalemb{\mathpalette\inner@dalemb{}}
\def\inner@dalemb#1{%
  \add@dalemb#1{03}%
  \add@dalemb#1{06}%
  \square
}
\def\add@dalemb#1#2{%
  \sbox0{\scalebox{1.#2}{$#1\square$}}%
  \rlap{\lower0.#2\ht0\box0}%
}
\makeatother

\makeindex

\DeclareMathAlphabet\mathbfcal{OMS}{cmsy}{b}{n}

\begin{document}
\thispagestyle{empty}
\bibliographystyle{elsarticle-harv} 



\begin{center}
\textcolor{black}{ \Large  \bf On a reformulation of Navier-Stokes equations based on Helmholtz-Hodge decomposition } \\
\vspace{3.mm}
{\bf Jean-Paul Caltagirone } \\
\textcolor{black}{\texttt{ \texttt{  \small calta@ipb.fr }  } } \\
\vspace{3.mm}
{ \small Bordeaux INP, University of Bordeaux, CNRS UMR-5295, \\
        Arts et Metiers Institute of Technology, INRAE, I2M Bordeaux, \\
        33400 Talence -- France  } 
\end{center}

\small

\textcolor{black}{\bf Abstract}

The proposal for a new formulation of the Navier-Stokes equations is based on a Helmholtz-Hodge decomposition where all the terms corresponding to the physical phenomena are written as the sum of a divergence-free term and another curl-free term. These transformations are founded on the bases of discrete mechanics, an alternative approach to the mechanics of continuous media, where conservation of the acceleration on a segment replaces that of the momentum on a volume. The equation of motion thus becomes a law of conservation of total mechanical energy per volume unit where the conservation of mass is no longer necessarily an additional law. The new formulation of the Navier-Stokes equations recovers the properties of the discrete approach without altering those of its initial form; the solutions of the classical form are also those of the proposed formulation. Writing inertial terms in two components resulting from the Helmholtz-Hodge decomposition gives the equation of motion new properties when differential operators are applied to it directly.

\vspace{3.mm}
\textcolor{black}{\bf Keywords}

Navier-Stokes equations, Discrete Mechanics;  Helmholtz-Hodge Decomposition; Inertial flows 

\normalsize

\vspace{-3.mm}
\begin{verbatim}
_______________________________________________________________
\end{verbatim}
\vspace{-2.mm}
This article may be downloaded for personal use only. Any other use requires prior permission of the author and AIP Publishing. 
\vspace{1.mm}

J-P Caltagirone, On a reformulation of Navier-Stokes equations based on Helmholtz-Hodge decomposition, Phys. Fluids, 33, 063605, https://doi.org/10.1063/5.0053412, 2021.

\vspace{-6.mm}
\begin{verbatim}
______________________________________________________________
\end{verbatim}

\vspace{-5.mm}

\normalsize

\section{Introduction}

The Navier-Stokes equations in their primitive form or expressed in various derived formulations represent the main foundation in fluid mechanics. They come from the concept of continuous medium which gives it the properties of derivation at a point, of integration, of analysis, etc.
Although there are still unresolved questions as to the existence of a regular solution to the incompressible formulation of these equations for a three-dimensional space, the Navier-Stokes equations associated with the continuity equation form the reference physical model for fluid flows.
They are representative of compressible or non-compressible viscous flows, two-phase flows, turbulence, acoustic waves, shock waves, etc. Constitutive equations, state laws or rheological laws can be associated to tackle complex problems of industrial or environmental interest.

A recently developed method approaches derivation of the equation of motion by considering that it is the acceleration which is kept on a rectilinear path and not the quantity of motion; conservation of mechanical energy is preferred over the conservation of mass which is set aside in the description of the movement. The fundamental law of dynamics is reinterpreted by considering that the proper acceleration of a material medium or of a particle with or without mass is the sum of the accelerations applied to it. The derived equation of motion becomes a law of conservation of energy per unit mass, or a sum of accelerations. The extension to two or three dimensions of space is carried out within a local frame of reference made up of a collection of segments, where the interactions between them are cause and effect. The notions of continuous medium, of Galilean or inertial frame of reference, of representation at a point of all quantities, are thus abandoned in favor of a geometric assembly made up of entangled structures referred to as primal and dual. The intrinsic acceleration of the material medium on a segment is described as the superposition of a direct acceleration carried by the primal structure and an induced acceleration carried by the dual structure. This intrinsic acceleration is thus written as a Helmholtz-Hodge decomposition, in a sum of a gradient of scalar potential and the curl of a vector potential \cite{Cal20a}. At the same time, the inertia terms are written in the form of a Helmholtz-Hodge decomposition of the inertial potential \cite{Cal20c}.
These equations are as representative of the flow of fluids as of the movements of solids or the propagation of waves in material media or a vacuum. Numerous simulations based on this discrete equation of motion \cite{Cal21b} have verified that the exact solutions obtained are also those of the Navier-Stokes equations, and have validated the results on classical test cases from the literature.

The objective here is to establish a new formulation of the Navier-Stokes equations on the basis of the properties of the equation of motion in discrete mechanics. Several choices required by the concept of continuous medium inhibit the possibility of formally transforming the Navier-Stokes equations into a Helmholtz-Hodge decomposition. First, the choice of momentum as the main variable, as the association of the variable mass in space and velocity is incompatible with such a decomposition. The physical properties not attached to a particular operator also lead to spatial derivations which cannot, in general, be written as the sum of a gradient and a curl. The classical inertia terms, in particular the Lamb vector, are not curls and the Helmholtz-Hodge decomposition of the inertia is also compromised.
To preserve the primal properties of the Navier-Stokes equations while incorporating some from discrete mechanics, all the terms of the classical form are transformed, (i) the viscous term is transformed into a rotational form where the kinematic viscosity is contained within the dual curl, (ii) the term containing compressive viscosity, an unclear quantity for fluids, is integrated into the mass conservation stress and (ii) the inertial term of Lamb is replaced by a term with free divergence.
Significant formal differences appear in the second order when the divergence and curl operators are applied to classical Navier-Stokes equations; they lead to additional terms whose existence is discussed. However, the analytical solutions and simulations from the Navier-Stokes equations are the same as for the new form. The transformations of the classical formulation of the equation of motion for fluids resulting from discrete mechanics have thus made it possible to endow it with new properties.

The first part of this article aims to recall the concepts of discrete mechanics and its main properties in an autonomous form. The primal and dual geometric structures are defined and the variables located within these structures; the derivation of the discrete equation of motion is then briefly presented. The formulation of the Navier-Stokes equations is then modified without changing the meaning as a function of the properties of the discrete equation. Finally, a standard test case used in continuum mechanics, that of stationary flow around a solid cylinder fixed at a Reynolds number $Re =  30$, is used in order to show the equivalence between the two classical and modified formulations of the Navier-Stokes equations.

\section{Discrete mechanics framework} 

The physical model is based on the concept of discrete mechanics \cite{Cal19a}, where the equation of motion is derived on an oriented segment taking into account all the effects of compression, viscosity, inertia, etc. This originally discrete approach becomes a continuous version that considers that the length of the segment tends towards zero; however, it has an attractive discrete interpretation when used to simulate the movement of fluids and solids on polygonal or polyhedral meshes.

The classical notions of inertial global frame of reference and continuous medium are replaced by those of local frame of reference and discrete medium. After a brief presentation of the geometric framework, the equations of discrete mechanics will be derived and their properties analyzed.

\subsection{Geometric representation of local frame of reference}

\begin{figure}[!ht]
\begin{center}
\includegraphics[width=7.cm]{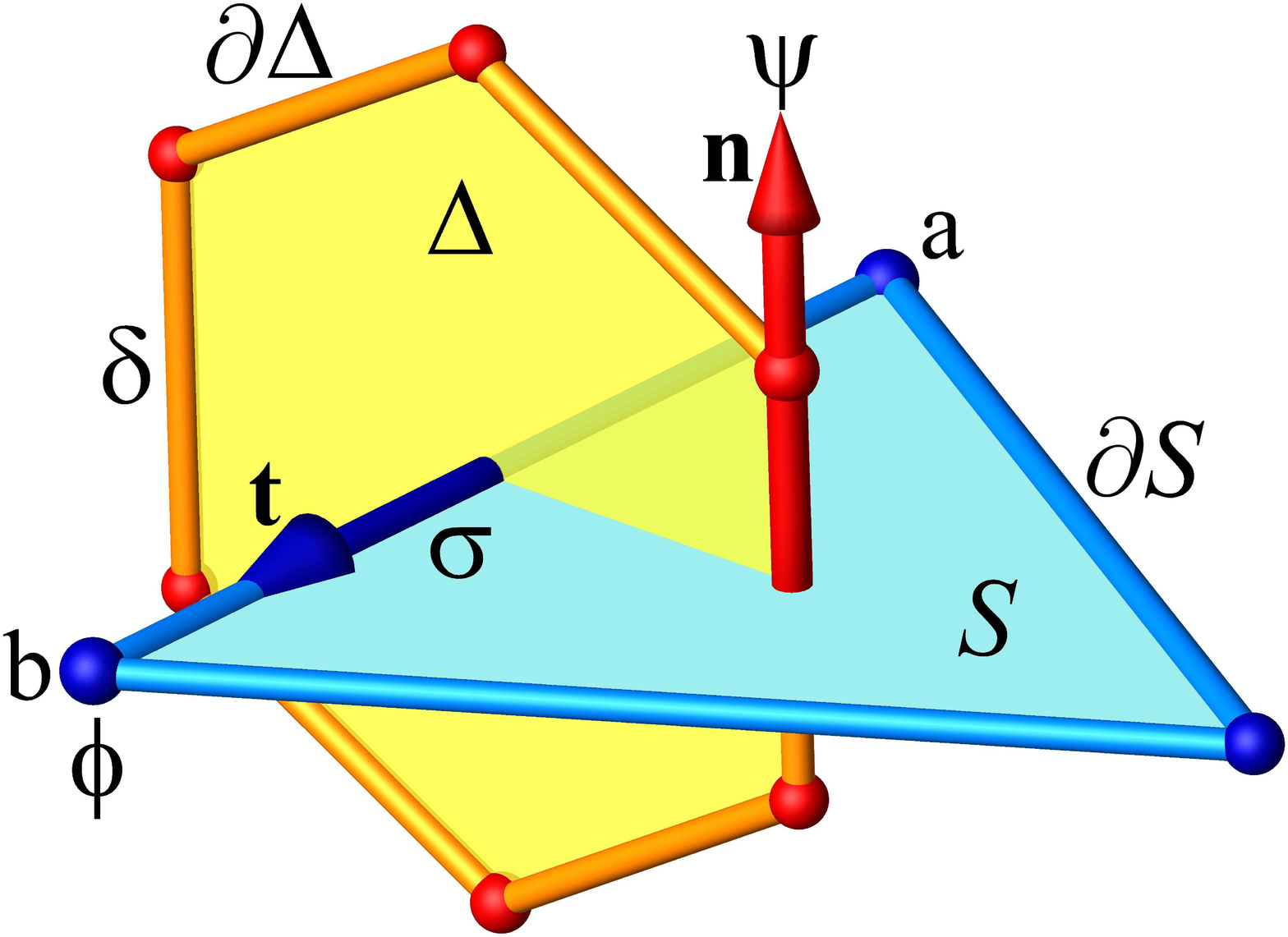}
\caption{ Geometric structure of discrete mechanics. The primal surface $S$ is defined by the edges $\sigma$ whose collection forms the boundary $\partial S$ of the surface; each edge oriented by the unit vector $\mathbf t$ is delimited by two points $a$ and $b$. Similarly, the dual surface $\Delta$ is composed by segments $\delta$ which are associated with the barycenters of each of the facets $S$ and oriented by the unit vector $\mathbf n$ and whose collection forms the boundary $\partial \Delta$. The unit vectors are orthogonal by definition, $\mathbf n \cdot \mathbf t = 0$. Besides, the direct orientation of circulation along $\partial S$ or $\partial \Delta$ is in agreement with the Maxwell's corkscrew rule. }
\label{primdual}
\end{center}
\end{figure}

Consider the geometric structure in figure (\ref{primdual}). It is made up of two families of sub-structures named primal and dual structures; the supposedly planar faces of primal structure $S $ are delimited by their edges $ \partial S $ and the faces of dual structure $\Delta$, limited by the edges $\partial \Delta$. Faces $S$ and $\Delta$ are assumed to be orthogonal by construction; their respective normals $\mathbf t$ and $\mathbf n$ are thus orthogonal, $\mathbf t \cdot \mathbf n = 0$. Limits $\partial S$ and $\partial \Delta$ are respectively composed of the oriented segments $\sigma$ and $\delta$. The orientations of segments $\sigma$ and $\delta$ follow Maxwell's corkscrew rule.

The variables and the physical properties are located in a unique way within the stencil of figure (\ref{primdual}). In general, the vectors in space are not defined; only their projections on segments $ \sigma $ and $ \delta $ will be retained as scalars on these oriented segments. So acceleration $ \bm \gamma $ will be scalar $ \gamma $ on segments $ \sigma $ of the primal structure, $ \bm \gamma = \gamma \: \mathbf t $. Likewise, velocity $\bm v$ will only represent the velocity defined in this edge-based formulation. The scalars will be defined at the vertices, $ a $ or $ b $ of the primal structure, such as for example $ \phi $, the scalar potential of the acceleration. The vector potential, denoted $ \bm \psi $, is a pseudo-vector in the sense of the mechanics of continuous media; here it is the vector carried by normal $ \mathbf n $ to surface $S $, $\bm \psi = \psi \: \mathbf n$, where $\psi$ is the scalar defined at the barycenter of the primal facet.

The physical properties will also be affected on precise positions of the stencil corresponding to the local frame of reference. The longitudinal celerity $c_l$ of the medium is located on the vertices of the primal structure, while transverse celerity $c_t$ or kinematic viscosity $\nu$ is defined on the barycenters of the facets $S$. Discrete mechanics requires the definition of different operators to establish communication from the primal structure to the dual structure, and vice versa. They are the classical operators of the continuous medium, gradient, divergence, curl, but with a meaning that is specific to discrete mechanics.

The discrete gradient of scalar $ \phi $ is the projection on $ \sigma $ of the vector of the space $ \nabla \phi $ but will be denoted with the same symbol for the sake of simplicity; so $ \nabla \phi $ will be the vector $ | \nabla \phi | \: \mathbf t$ assigned to the unit vector $ \mathbf t $. Vector $ \nabla \times \bm v $ is the primal curl of velocity $\bm v$ carried by $ \mathbf n $. This curl is calculated from Stokes' theorem as the circulation along contour $ \partial S $ of vectors $ \bm v $. Similarly, the dual curl, denoted $ \nabla ^ d \times \bm \psi $, is the line integral over $ \partial \Delta $ of vector $ \bm \psi $. Finally, the divergence of the velocity, denoted $\nabla \cdot \bm v$, is the flux of the projections of this vector $ \nabla \cdot \bm v_k | A_k | $ affected by the areas of the corresponding dual facets $ | A_k | $ and divided by the dual volume $ | D | $. These operators discretely mimic certain properties of continuous operators, in particular $ \nabla \cdot (\nabla ^ d \times \bm \psi) = 0 $ and $ \nabla \times (\nabla \phi) = 0 $, whatever the polygonal or polyhedral structures considered \cite{Lip14}. Moreover, \cite{Cal19a} we show the orthogonality of gradient and curl operators $ \nabla \phi \: \cdot \nabla ^ d \times \bm \psi = 0 $.

Figure (1) corresponds to the minimal structure which allows construction of complex geometries, objects or domains in two or three dimensions of space. The primal and dual faces are composed of polygons or polyhedra with any number of faces. The primal structures of the structured or unstructured finite element are generated by the classically available mesh generators. When the mesh is of Cartesian structured type, we find the formulation used for the Marker And Cell \cite {Har65} scheme that is generalized by this methodology.

The subject exposed in this context concerns the physical models of mechanics rather than the methodologies which serve to solve the associated equations. However, these two fields of simulation have come closer with the development of methods resulting from differential geometry.

The Discrete Exterior Calculus (DEC) \cite{Hir03, Des05, Hir16, Ham18} method is based on differential geometry and exterior calculus, where primal and dual structures are used to apply operators (exterior derivative, Hodge operator). Each term of the chosen continuous equation is then replaced by its discrete version. The operators mimic some of the important discrete properties of the continuous medium, notably $\nabla \times \nabla \cdot \phi = 0$ and $\nabla \cdot \nabla \times \bm \psi = 0$ for all regular functions $\phi$ and $\bm \psi$. One advantage of discretizing with DEC is that, by construction, Stokes' theorem is always verified exactly as in discrete mechanics. At the same time, the mimetic methods resulting from the pioneering work of Shaskov \cite {Hym99, Hym99b} have been applied successfully to many equations in physics, such as the diffusion equation, Darcy's law, Stokes, Euler, Navier-Stokes, Navier-Lam{\'e}, or Maxwell. Overall, the formalism associated with these techniques is close to that of DEC. Other discretization methods result from the first works on DEC and on the mimetic method of \cite{Per00, Pal17, Ded19, Bon14, Bon15, Bon15b}. The differences with the previous methods often relate to the nature of the discretization and the location of the unknowns in the stencil of the geometric structure.

The numerical methodology resulting from discrete mechanics is clearly part of the preceding methods, but it is distinguished by the physical model. It is above all a reformulation of the equations of mechanics which departs from the framework of the mechanics of continuous media by associating the physical model with the geometrical structure in the derivation of the equation of motion.

\subsection{Discrete mechanics equation of motion} 

The physical model developed in recent years is based on the conservation of acceleration on segment $ \sigma $ of the primal structure shown in figure (\ref{primdual}). This choice stems from Galieo's weak equivalence principle (WEP) which mentions that the mass related to inertia is equal to the mass due to gravitation. This principle is now verified at less than one part for $10^{15}$. Newton's second law translates this phenomenon into modern language by establishing the link between force and the product of mass by acceleration, $ \mathbf F = m \: \bm \gamma $ or $ m \: \mathbf g = m \: \bm \gamma $ for the gravitational force alone. Discrete mechanics returns to this interpretation by writing the equality $ \bm \gamma = \mathbf g $ and, more generally, $ \bm \gamma = \bm h $ where $ \bm \gamma $ is the intrinsic acceleration of the material medium or of a particle and $ \bm h $ the sum of all the external accelerations applied to this medium. The mass disappears from the equation of motion, which will become a law of conservation of energy. As the theory of relativity establishes the link between mass and energy, the conservation of mass will become an implicit law of conservation.

The notions of continuous medium, components and global inertial frame of reference are abandoned in favor of a local frame of reference based on the structure in figure (\ref{primdual}). The vectors of space are replaced by their projections on oriented segments. An important constraint will, however, have to be respected by the equation of discrete movement, which will have to filter the rigid uniform movements of translation and rotation. The material medium or the particle in one of these movements or in any linear combination will not have to undergo any action which allows it to change its trajectory.

Like any vector, acceleration can be split into a gradient of a scalar potential, denoted $ \phi $, and the curl of a vector potential, denoted $ \bm \psi $, as a Helmholtz-Hodge decomposition. The presence of a harmonic term, both curl-free and divergence-free, is zero because it is directly associated with uniform movements of translation and rotation. The equation of motion reads:
\begin{eqnarray}
\displaystyle{ \bm \gamma = - \nabla \phi + \nabla^d \times \bm \psi } 
\label{newton}
\end{eqnarray}
where $\nabla^d \times \bm \psi$ is the dual curl of the vector potential of the acceleration.

From the physical point of view, the intrinsic acceleration on a segment $ \sigma $ can only be due to two well-defined causes: (i) a direct acceleration defined by the gradient of potential $ \phi $ between the two ends $ a $ and $ b $, and (ii) an acceleration induced by the circulation of vector $ \bm v $ on all the segments of the different primal facets having segment $ \sigma $ in common; it is characterized by the dual curl of $\bm \psi $. This is what equation (\ref{newton}) translates; the analogy with electric and magnetic fields is obvious. The quantities $ | \bm \gamma | $ and $ | \bm v | $ are considered as constants on segment $ \sigma $ and the first integral equation of motion is written:
\begin{eqnarray}
\displaystyle{  \int_{\sigma} \bm \gamma \cdot \mathbf t \: dl = - \int_{\sigma} \nabla  \phi  \cdot \mathbf t \:  dl + \int_{\sigma} \nabla^d \times \bm  \psi \cdot \mathbf t \:  dl  } 
\label{bilanlin}
\end{eqnarray}

The integral of the proper acceleration between $ a $ and $ b $ represents the total variation per unit mass between these two points, the sum of kinetic energy and potential energy:
\begin{eqnarray}
\displaystyle{  \Phi_b - \Phi_a = \int_a^b \bm \gamma \cdot \mathbf t \: dl   } 
\label{energy}
\end{eqnarray}
where $\Phi$ is total energy per mass unit. Thus the equation of motion will in fact be a law of conservation of energy on $ \sigma $.

The absence of a global frame of reference poses the problem of extension to several dimensions of space, and the no less important problem of changing the frame of reference. It will be necessary to forget recourse to this latter operation, as it will not be possible to use this important property of the notion of continuous medium extended by special relativity to Lorentz transformations. The length of segment $ [a, b] $ defines what we can call the discrete horizon $ dh = [a, b] $, linked to a time lapse $ dt $ by the celerity of the medium $c_l$ by $dh = dt \: c_l$. No information can be propagated beyond this distance for the period of time considered. For longer times, the information will be transmitted in a cause and effect relation from one local frame of reference to another through the variation of the scalar potential located at the common vertex of these two frames of reference. The extension to space dimensions larger than unity will be carried out according to this principle of proximity.

The concepts developed above make it possible to write a law of motion for any dimension of space for vector $ \bm \gamma $: 
\begin{eqnarray}
\displaystyle{  \bm \gamma = \frac{d \bm v}{d t}  = -  \nabla  \left( \phi^o + d \phi \right) +  \nabla^d \times \left( \bm  \psi^o + d \bm \psi \right)  } 
\label{loidynhh}
\end{eqnarray}
where $\phi^o$ and $\bm \psi^o$ are the retarded potentials with reference to the Li{\'e}nard and Wichert potentials \cite{Lie98} at instant $t^o$ and $\phi = \phi^o + d \phi$ and $\bm \psi = \bm \psi^o + d \bm \psi$ the actual potentials at time $t = t^o + dt$. This very important notion of discrete mechanics corresponds to the accumulation of the compression and rotation energies attached respectively to the terms $\nabla \phi$  and $\nabla^d \times \bm \psi$. The history of the material medium is thus characterized by potentials $\phi^o$ and $\bm \psi^o$ since an initial moment $t^o = 0$, a known state, where their values are equal to $\phi_0$ and $\bm \psi_0$.

Variations in potentials $d \phi$ and $d \bm \psi$ are modeled by referring to the simple compression and shear experiments with which these two potentials are respectively associated. The first corresponds to an experiment of isothermal compression of a fluid in a cavity whose volume is reduced at constant velocity by a piston for a time $dt$; the pressure required is a function of the compressibility of the fluid and therefore of its longitudinal celerity. The second experiment corresponds to shearing of a fluid at uniform velocity and the force to be exerted is a function of the transverse celerity.
In a trivial way, the pressure increment is a function of the divergence of the injection velocity for compression and the shear stress is a function of the curl of velocity. Analysis of these experiments first leads to the integral form of the equation of motion:
\begin{eqnarray}
\displaystyle{  \int_{\sigma} \bm \gamma \cdot \mathbf t \: dl = - \int_{\sigma} \nabla \left( \phi^o - \: dt \: c_l^2 \: \nabla \cdot \bm v \right) \cdot \mathbf t \:  dl + \int_{\sigma} \nabla^d \times \left(\bm  \psi^o - dt \: c_t^2 \: \nabla \times \bm v \right) \cdot \mathbf t \:  dl  } 
\label{disurG2}
\end{eqnarray}

The update of potentials $\phi^o$ and $\bm \psi^o$ over time corresponds to a process of accumulation of compression and shear energies which makes the discrete equation a continuous memory model. Contrary to the classical equations which describe an instantaneous equilibrium, the discrete equation introduces a memory into the equation at the origin of the exchanges between the potential and kinetic energies. These integrals transform the differential equation into an integro-differential equation, which complicates its mathematical analysis and its solution. At each instant the equation represents a mechanical equilibrium and the passage from an instant $ t^o$ to instant $t$ is carried out by its resolution.

The retarded potentials represent the energies accumulated between $ t = 0 $ and $ t ^ o $:
\begin{eqnarray}
\displaystyle{   \phi^o = - \int_0^{t^o} \: c_l^2 \: \nabla \cdot \bm v \:  d\tau;  \:\:\:\:\:\:\:\:\:\:
 \bm \psi^o = - \int_0^{t^o} \:  c_t^2  \: \nabla \times \bm v \: d\tau } 
\label{disint}
\end{eqnarray}

It is necessary to consider the part of the energies really accumulated over time. In fact, the longitudinal and transverse waves can be attenuated over time; for example, transverse waves in Newtonian fluids are attenuated over time periods of the order of $10^{-12} \: s$, while elastic media conserve energies entirely. Two physical quantities are defined to translate the reduction in energy over time: longitudinal attenuation factor $ \alpha_l $ and transverse attenuation factor $ \alpha_t $. These quantities, between $ 0 $ and $ 1 $, can be functions of time and space through other variables. In the case of a purely viscous Newtonian fluid, the attenuation factor $\alpha_t$ is equal to the unity and quantity $ dt \: c_t^2$ must be replaced by kinematic viscosity $\nu$.

The discrete equation of motion then becomes:
\begin{eqnarray}
\left\{
\begin{array}{llllll}
\displaystyle{ \bm \gamma = - \nabla  \left( \phi^o - dt \: c_l^2 \: \nabla \cdot \bm v \right) + \nabla^d  \times   \left( \bm \psi^o - dt \: c_t^2 \: \nabla \times \bm v \right) - \kappa \: \bm v } \\  \\
\displaystyle{ \phi  = ( 1 - \alpha_l ) \: \phi^o -  dt \: c_l^2 \: \nabla \cdot \bm v   } \\ \\
\displaystyle{ \bm \psi = ( 1 -  \alpha_t ) \:  \bm \psi^o -  dt \: c_t^2 \: \nabla \times \bm v    }
\end{array}
\right.
\label{discrete}
\end{eqnarray}

The term $- \kappa \: \bm v$ is added to the velocity equation; it expresses the introduction of a drag applied on segment $ \sigma $ which, like the other terms of the equation, is expressed in the form of an acceleration; $ \kappa $ is a quantity whose unit is given in $m^2 \: s^{-2}$, that is, an energy per unit of mass, the energy transformed into heat on the segment. This term is similar to the one present in Darcy's law, $- \nabla p - \mu / K \: \bm v$, where $K$ is the permeability of the porous medium; in discrete mechanics this equation becomes $ - \nabla \phi - \kappa \: \bm v $ where $ \kappa = \nu / K$.

In continuum mechanics, the material derivative can be written indifferently in the forms $\bm v \cdot \nabla \bm v $, $ \nabla \cdot \left (\bm v \otimes \bm v \right) - \bm v \: \nabla \cdot \bm v$ or even in a rotational form $\nabla (| \bm v |^2/2) - \bm v \times \nabla \times \bm v $; the last term of this form is called the Lamb vector. The divergence of this term is not zero and reveals two terms, flexion and enstrophy, representing in a continuous medium certain properties of turbulent flows \cite{Ham08}. Note that this Lamb term is not a curl term. In discrete mechanics the form of the terms of inertia is different; the reasons for this modification are given in \cite{Cal20a} and \cite{Cal20c}, which is presented as a Helmholtz-Hodge decomposition of an inertial potential $ \phi_i$.

Thus the inertial potential $\phi_i = | \bm v |^2/2$ will give a physical meaning to the material derivative:
\begin{eqnarray}
\displaystyle{ \bm \gamma  \equiv \frac{d \bm v}{dt} = \frac{\partial \bm v}{\partial t} + \nabla \left( \frac{| \bm v |^2}{2} \right) - \nabla^d \times \left( \frac{| \bm v |^2}{2} \mathbf n \right) }
\label{material}
\end{eqnarray}

The scalar potential $ \phi_i $ is defined in the whole space, but particularly at the vertices and at the barycenters of the facets $ S $ of the primal structure. The dual curl of vector $\phi_i \: \mathbf n$ projects an acceleration on $\sigma$ which competes with that set by $ \nabla \phi_i $. If the trajectory of the material medium or of the particle is rectilinear, it is because the two terms are compensated for; this is the case of the plane Couette and Poiseuille flows for example, while each of the terms is not zero. In a way, the terms of inertia represent the curvature of the inertial potential.

The specific form of the terms of inertia in discrete mechanics re-establishes the symmetry of all the terms of the equation of motion: all the terms are either divergence-free or curl-free; this confers interesting properties when operators, such as divergence or primal curl, are applied directly to the equation of motion \cite{Cal20a}.

\vspace{2.mm}

The simultaneous processing of movements in media of different natures in monolithic mode, for example a fluid and a solid in Fluid-Structure Interaction, can be achieved using the Darcy term $ - \kappa \: \bm v $. This penalization technique \cite{Ang12} allows simulations to be carried out very simply where the velocity becomes zero in the solid parts of the flow by fixing $ \kappa \rightarrow \infty $. As potential $ \phi $ at the interfaces is perfectly defined in the fluid, it is possible to satisfy the equation $- \nabla \phi - \kappa \: \bm v = 0$ in the solid with $\bm v \rightarrow 0$. The calculation of the solution $ (\phi, \bm \psi, \bm v) $ can be carried out simultaneously in the two media or separately by using the potentials at the interfaces of the fluid to then calculate the solution in the solid.

In fact Darcy's law can be rewritten in terms of vector potential, the two forms are then:
\begin{eqnarray}
\left\{
\begin{array}{llllll}
\displaystyle{ - \nabla \phi - \kappa \: \bm v = 0  } \\ \\
\displaystyle{  \nabla^d \times \bm \psi - \kappa \: \bm v = 0   }
\end{array}
\right.
\label{darcy}
\end{eqnarray}

Indeed, in classical mechanics, flows in non-inertial irrotational porous media can be apprehended from the scalar potential alone, but nothing prevents us from writing that the flow in a porous medium derives from a vector potential $ \bm \psi $, such that  $- \nabla \phi = \nabla ^ d \times \bm \psi$. If the flow is incompressible, the divergence of the first equation of (\ref{darcy}) and the curl of the second equation give $\nabla^2 \phi = 0$ and $\nabla^2 \bm \psi = 0$ or, in two-dimensional space $\nabla^2 \psi = 0$, with $\psi = \bm \psi \cdot \mathbf n$.

\subsection{Properties of the formulation}

The discrete formulation and the equation (\ref{discrete}) have specific properties, in particular due to the abandonment of the notion of continuous medium; this generates difficulties such as the inability to change the frame of reference but presents remarkable differences, mainly due to the Helmholtz-Hodge decomposition, especially when differential operators are applied to the equation. 
For all polygonal or polyhedral geometric structures, certain properties of the continuous medium are mimicked by the discrete; for example, we can show \cite{Cal20a} that $ \nabla \phi$ and $\nabla^d \times \bm \psi$ are globally and locally orthogonal, $\nabla \phi \cdot \nabla^d \times \bm \psi = 0$.

Two other important properties of continuum mechanics, $\nabla \times \nabla \phi = 0 $ and $\nabla \cdot \nabla \times \bm \psi = 0$, are indispensable in discrete mechanics.  It is easy to show, on the primal structure, that the discrete curl of a discrete gradient is zero:
\begin{eqnarray}
  \left\{ 
\begin{array}{llllll}
\displaystyle{   \int_a^b  \nabla \phi  \cdot \mathbf t \: dl = \phi_b - \phi_a } \\  \\
\displaystyle{  \int_{\partial S}  \nabla \phi  \cdot \mathbf t \: dl  = 0 } \\  \\
\displaystyle{  \int \!\!\!\! \int_{ S} \nabla \times \big( \nabla \phi \big) \cdot \mathbf n \: ds  = 0 } \\  \\
\displaystyle{  \nabla \times \big( \nabla \: \phi \big)  = 0 }
\end{array}
\right.
\hspace{10.mm}
  \left\{ 
\begin{array}{llllll}
\displaystyle{  \sum_{i=1}^n \:  \int \!\!\!\! \int_{S} \nabla^d \times \bm \psi \cdot \mathbf n \: ds = 0 } \\  \\
\displaystyle{  \int \!\!\!\! \int_{ S} \big( \nabla^d \times \bm \psi \big) \cdot \mathbf n \: ds  = 0 } \\  \\
\displaystyle{  \int \!\!\!\! \int \!\!\!\! \int_{ D} \nabla \cdot \big( \nabla^d \times \bm \psi \big) \: ds  = 0 } \\  \\
\displaystyle{  \nabla \cdot \big( \nabla^d \times \bm \psi \big)  = 0 }
\end{array}
\right.
\label{propriop}
\end{eqnarray}

\begin{figure}[!ht]
\begin{center}
\includegraphics[width=6.cm]{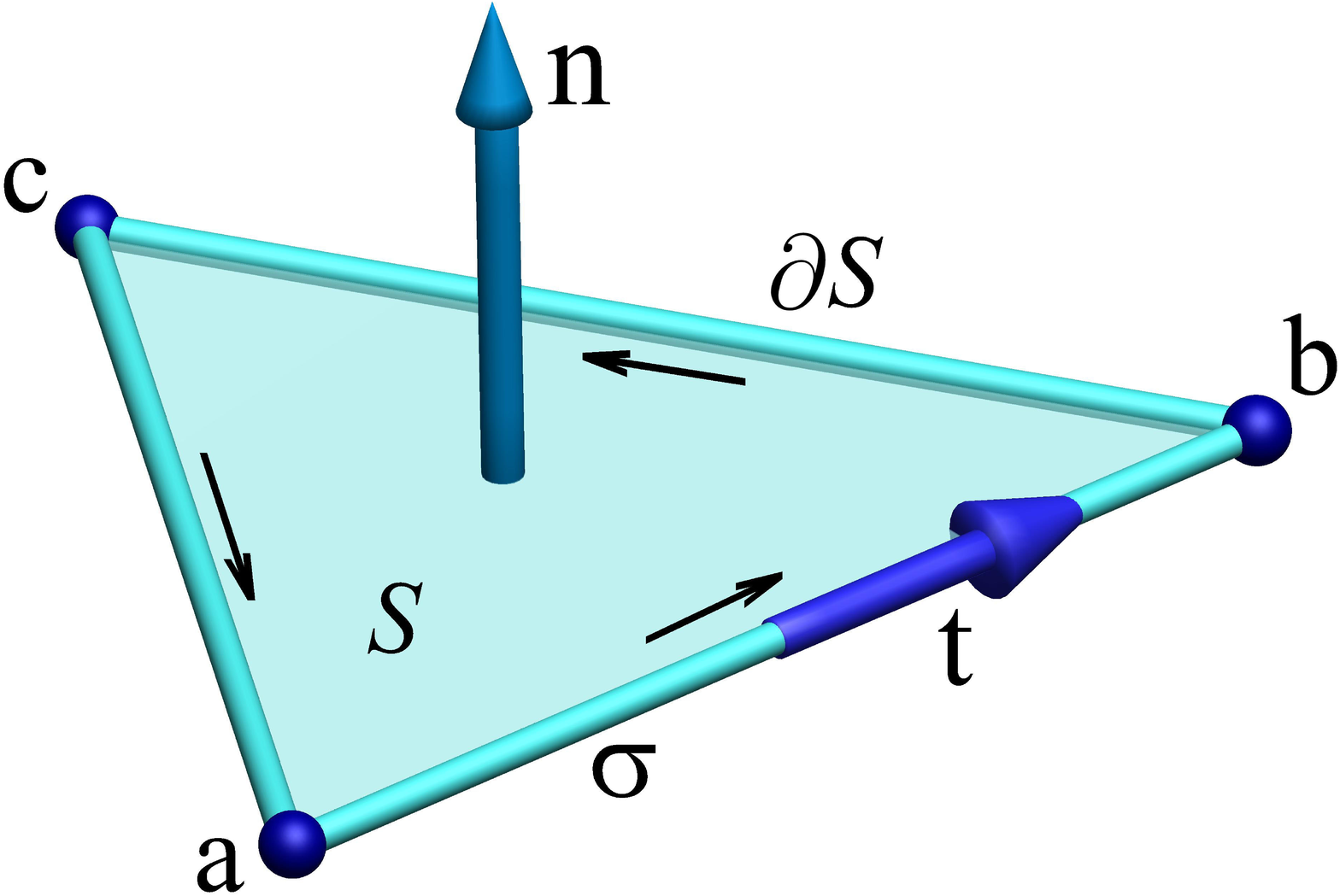}
\hspace{10.mm}
\includegraphics[width=5.cm]{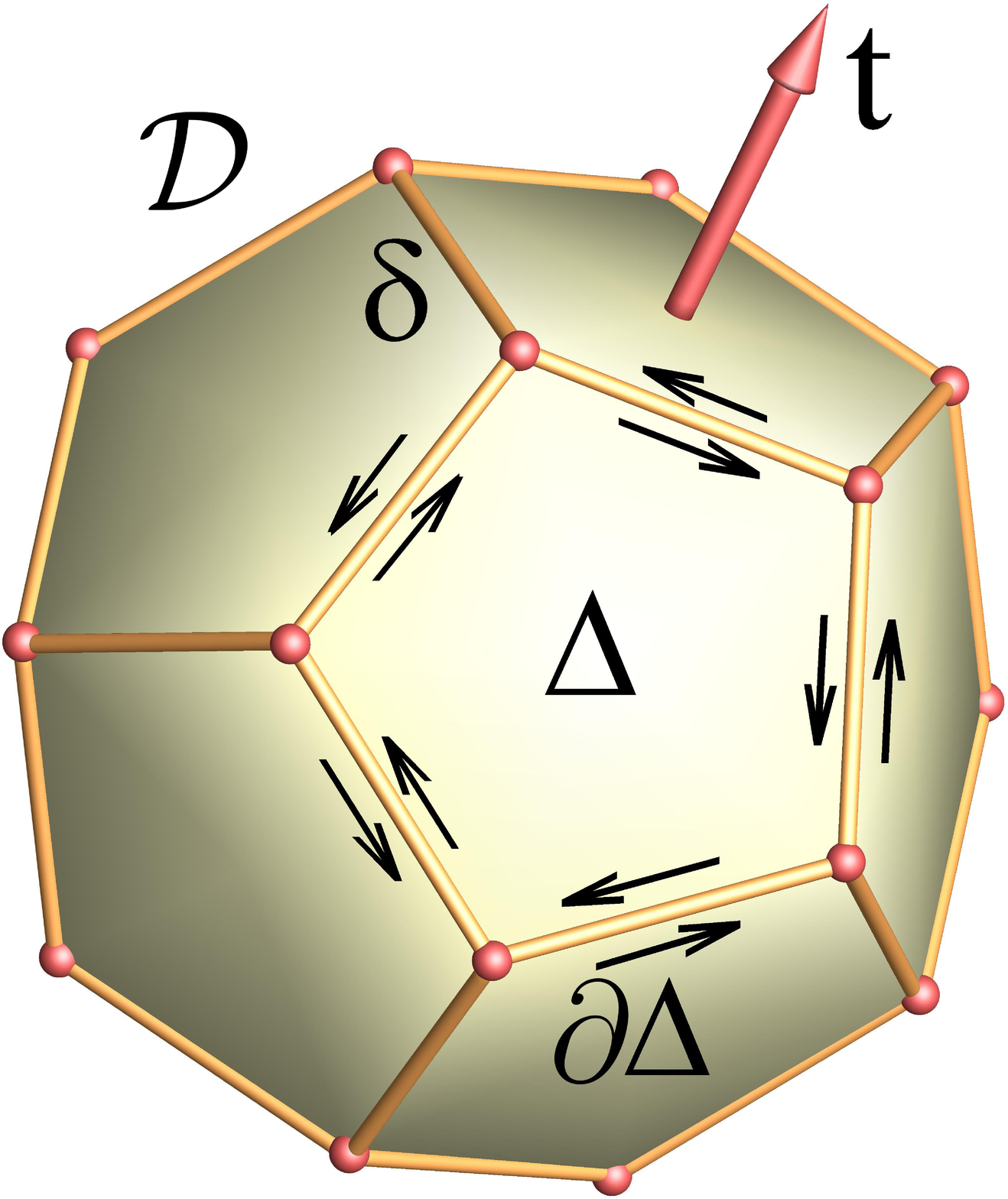}
\vspace{-1.mm}
\caption{ Primal structure at left where $\nabla \times \nabla \phi= 0$ and, at right, $\nabla \cdot \nabla^d \times \bm \psi= 0$. }
\label{propri}
\end{center}
\end{figure}

Similarly, the discrete divergence of the discrete primal curl calculated on the dual volume is zero.
Figure (\ref{propri}) shows how the property $\nabla \times \nabla \phi = 0$ is evaluated on the primal structure and how $\nabla \cdot \nabla^d \times \bm \psi = 0$ is verified on the dual structure.

Figure (\ref{propri}) shows that the condition for obtaining the property $ \nabla \times \nabla \phi = 0 $ is that the contour $ \partial S $ is closed and that the second $ \nabla \cdot \nabla ^ d \times \bm \psi = 0 $ is obtained if the segments forming the dual volume closed by dual facets $ \Delta $ are traversed twice in the opposite direction. This amounts to considering that the integral of vector potential $ \bm \psi $ on all the facets of the dual volume is equal to zero.

\vspace{1.mm}

The classical rules used to obtain identities in vector calculus are sometimes unsuitable for defining certain compositions of operators. This is because the primal curl and the dual curl are not differentiated within the framework of the notion of continuous medium. Thus the fourth identity in Table (\ref{calculprop}) has a meaning in vector calculus, while the result is not defined in discrete mechanics and must be replaced by the third identity where the external operator is a dual curl. Table (\ref{calculprop}) gives the few definitions useful for the transformation of the equations treated here.

\begin{table}[!ht]
\begin{center}
\small
\begin{tabular}{|c|c|c|c|c|}   \hline
    Second derivatives identities          &  Property             \\ \hline  \hline
 $ \nabla \cdot \nabla^d \times \bm \psi$  &       0      \\ \hline
 $ \nabla \times \nabla \phi$              &       0      \\ \hline
 $ \nabla^d \times \nabla \times \bm v$    &   $\nabla \left( \nabla \cdot \bm v \right) - \nabla^2  \bm v$ \\ \hline
 $ \nabla \times \nabla \times \bm v$      &       undefined      \\ \hline
 $ \nabla \cdot \nabla \cdot \bm v$        &       undefined      \\ \hline
 $ \nabla \times \nabla \cdot \bm v$       &       undefined      \\ \hline
\end{tabular}
\normalsize
\caption{Some properties of operators where $ \bm v $ is a true vector or polar vector, $ \bm \psi $ a pseudo-vector or axial vector and $ \phi $ a scalar.  }
\label{calculprop}
\end{center}
\end{table}

\vspace{1.mm}
The discrete equation is completely formulated without tensor, including of order one, if we consider that vector $ \bm v $ is scalar $ | \bm v | $ attached to oriented segment $ \sigma $. All discrete operators have a simple geometric meaning. This reduction in no way precludes consideration of mechanical properties which depend on direction; in this case the matrix corresponding to the tensor is diagonalized and the projection of the property is then assigned to segment $\sigma$.

Equation (\ref{discrete}) is entirely kinematic; the constitutive laws, rheological laws and state laws are absent from the discrete formulation. They only intervene through the definition of potentials $ \phi $ and $ \bm \psi $. For example, for an isentropic compressible flow of an ideal gas, the scalar potential will take the form $\phi = p \: \rho^{\gamma}$, where $p$ is the pressure, $\rho$ the density and $\gamma$ the ratio of specific heat at constant pressure and volume. This greatly facilitates the treatment of surface discontinuities or shock waves. The solution to these problems is always defined by the quantities $ (\phi, \bm \psi, \bm v) $. The density can be upgraded explicitly from the divergence of the velocity and the material derivative:
\begin{eqnarray}
\displaystyle{ \frac{d \rho}{d t} = - \rho \: \nabla \cdot \bm v }
\label{mass}
\end{eqnarray}

The density is then advected by the velocity field using the term $ \bm v \cdot \nabla \rho$ for diffuse fields or with an appropriate methodology (Volume Of Fluid, Front-Tracking, Arbitrary Lagrangian-Euleraian methods , etc.) for tracking interfaces.

Another scalar potential is introduced here which is not directly linked to the acceleration, denoted $ \varphi $; it can be defined at the same time on the vertices and on the barycenters of the primal structure. Vector $ \bm \varphi = \varphi \: \mathbf n $ will be orthogonal to facet $ S $. This potential may refer to different physical phenomena: gravitation $ \varphi_g $, capillary effects $ \varphi_c $, etc. The corresponding acceleration:
\begin{eqnarray}
\displaystyle{ \bm \gamma_s = - \nabla \varphi_s + \nabla^d \times \bm \varphi_s }
\label{source}
\end{eqnarray}
will represent the curvature of potential $ \varphi_s $. In continuum mechanics we say that a force derives from a potential when we can write $ \mathbf f = - \nabla  P $. In discrete mechanics any acceleration derives from two potentials, one scalar and the other vector. They are entangled for the intrinsic acceleration $ \bm \gamma = - \nabla \phi + \nabla ^ d \times \bm \psi $. In the other cases they come from the same scalar potential $ \varphi $. The intrinsic properties of the formulation are strictly verified for any polygonal and polyhedral structures. 

The physical model is a continuous memory one; prediction of the state of a physical system by an equation can in fact only be considered if its history is known from its initial state at time $t_0$. For example, an isolated observer whose movement is at constant velocity cannot detect whether it is in subsonic or supersonic regime; an equation describing an instantaneous state does not provide knowledge of certain quantities, such as the energy accumulated in previous states. Discrete equation (\ref{discrete}) includes the retarded potentials $\phi^o$ and $\bm \psi^o$ representing the energies per unit mass accumulated over time. These quantities make it possible to predict the response at the current moment according to the various imposed accelerations.

The integro-differential equation obtained naturally presents mathematical difficulties of analysis and resolution, but it ensures complete symmetry and conservation properties defined by Noether's theorem \cite{Noe11}.

Previous results \cite{Cal20a, Cal20c} on different test cases of fluid mechanics, solids and FSI have shown the effectiveness of the formulation. In the simplest cases the solutions correspond exactly to those of the equations of Stokes, Darcy, Navier-Stokes, Navier-Lam{\'e}, etc. Even if these simple solutions are identical, the fact remains that fundamental theoretical differences exist and make this formulation an alternative physical model which can be extended to other fields of physics, in particular electromagnetism.

\section{Transposition to Navier-Stokes equations} 

The objective is to obtain a formulation of the Navier-Stokes equations that preserves all of their properties and representativeness by adding certain issues from the equation of motion of discrete mechanics. The frame is now that of the continuous medium, where the vectors of $\mathcal R^3$ are represented by its components in an orthonormal frame of reference $(x, y, z)$.

The most classic form of the Navier-Stokes equation in its incompressible version is:
\begin{eqnarray}
\left\{
\begin{array}{llllll}
\displaystyle{  \rho \: \left(  \frac{\partial \bm v}{\partial t} + \bm v \cdot \nabla  \bm v \right)   =  - \nabla p + \nabla \left( \lambda \: \nabla \cdot \bm v \right) + \nabla \cdot \left( \mu \left( \nabla \bm v + \nabla^t  \bm v \right) \right) }  \\ \\
\displaystyle{ \nabla \cdot \bm v  =  0  }
\end{array}
\right.
\label{navier}
\end{eqnarray}
where $\bm v$ is, in this section, the velocity vector of space $\mathcal R^3$, $p$ the pressure, $\lambda$ and $\mu$ the viscosity coefficients of Lam{\'e}. Traditionally the Stokes law $3 \: \lambda + 2 \: \mu = 0$ is adopted in order to define the pressure as the trace of the tensor of the strain rates. The motion is here assumed to be incompressible for greater clarity, but this assumption does not limit the point developed for compressible flows where the mass conservation equation replaces $\nabla \cdot \bm v = 0$. Constitutive relations can optionally be added if necessary.

\subsection{Viscous and compression terms}

Let us first consider the viscous term which, in continuum mechanics, is written $\nabla \cdot \bm {\sigma}$, where $\bm {\sigma}$ is the strain rate tensor. Its divergence can be expressed as follows:
\begin{eqnarray}
\hspace{-2mm} \nonumber
 & \nabla \cdot \bm{\sigma} =\hspace{-1mm}  &   \nabla \cdot \left( - p \: \mathbf I + \lambda \: \nabla \cdot \bm v \: \mathbf I + 2 \: \mu \: \mathbf D \right) \\ \nonumber
\vspace{2.5mm}
\hspace{-2.mm} &\nabla \cdot \bm{\sigma} = &   \nabla \left(  - p  +  \lambda \: \nabla \cdot \bm v \right)  + 2 \: \mu \: \nabla \cdot \mathbf D + 2 \: \nabla \mu \cdot \mathbf D   \\ \nonumber
\vspace{2.5mm}
\hspace{-2.mm} &\nabla \cdot \bm{\sigma} = &   - \nabla p + \lambda \: \nabla ( \nabla \cdot \bm v ) + \nabla \cdot \bm v \: \nabla \lambda + \mu \nabla \cdot \left( \nabla \bm v + \nabla^t \bm v \right)  + \nabla \mu \cdot \left( \nabla \bm v + \nabla^t \bm v \right)  \\
\vspace{2.5mm}
\hspace{-2.mm}      &\nabla \cdot \bm{\sigma}  =  &     - \nabla p +  (\lambda + \mu) \:  \nabla (  \nabla \cdot \bm v ) + \mu \: \nabla^2 \bm v + \nabla \cdot \bm v \: \nabla \lambda + \nabla \mu \cdot \left( \nabla \bm v + \nabla^t \bm v \right)  \\   \nonumber
\vspace{2.5mm}
\hspace{-2.mm}       &\nabla \cdot \bm{\sigma}  =  &     - \nabla p  +  (\lambda + 2 \: \mu) \:  \nabla ( \nabla \cdot \bm v ) - \mu \: \nabla \times \nabla \times \bm v + \nabla \cdot \bm v \: \nabla \lambda + \nabla \mu \cdot \left( \nabla \bm v + \nabla^t \bm v \right) \\ \nonumber
\vspace{2.5mm}
 \hspace{-2.mm}     &\nabla \cdot \bm{\sigma}  =  &     - \nabla \Big( p  -  (\lambda + 2 \: \mu) \:   \nabla \cdot \bm v  \Big) - \mu \: \nabla \times \nabla \times \bm v + \nabla \cdot \bm v \: \nabla \lambda + \nabla \mu \cdot \left( \nabla \bm v + \nabla^t \bm v \right).  \nonumber
\label{tensdef}
\end{eqnarray}

where $\mathbf I$ is the unit tensor and $\mathbf D = (\nabla \bm v + \nabla^t \bm v ) / 2$ is the strain rate tensor.

Taking this expression into account, the equation of motion is then rewritten in the form:
\begin{eqnarray}
 \left\{  
\begin{array}{llllll}
\displaystyle{ \rho \:  \frac{d  \bm v }{d t  }   =   - \nabla \left(   p  -   (\lambda + 2 \: \mu) \:  \nabla \cdot \bm v  \right) - \mu \: \nabla \times \nabla \times \bm v  
 + \nabla \cdot \bm v \: \nabla \lambda + \nabla \mu \cdot \left( \nabla \bm v + \nabla^t \bm v \right)   }  \\  \\
\displaystyle{ \nabla \cdot \bm v  =  0  }
\end{array}
\right.
\label{NSrot}
\end{eqnarray}

At this stage no modification is made to the original version (\ref{navier}). Consider momentarily the viscosity coefficients of Lam{\'e} $\lambda$ and $\mu$ are constants (or piecewise constants). Likewise, consider that density $\rho$ is constant. With these provisional assumptions we have:
\begin{eqnarray}
 \left\{  
\begin{array}{llllll}
\displaystyle{  \frac{d \bm v}{dt} =   -\nabla \left( \frac{p}{\rho} -  \frac{(\lambda + 2 \: \mu)}{\rho} \:   \nabla \cdot \bm v  \right) - \frac{\mu}{\rho} \: \nabla \times \nabla \times \bm v  } \\  \\
\displaystyle{ \nabla \cdot \bm v  =  0  }
\end{array} 
\right.
\label{NSrot2}
\end{eqnarray}

By setting $\eta = (\lambda + 2 \: \mu) / \rho$ and $\nu = \mu / \rho$ and by expressing the material derivative, the vector equation (\ref{NSrot2}) becomes:
\begin{eqnarray}
\displaystyle{  \frac{\partial \bm v}{\partial t} + \frac{1}{2} \: \nabla \left( |\bm v |^2 \right) - \bm v \times \nabla \times \bm v =   -\nabla \left( \frac{p}{\rho} -  \eta \:   \nabla \cdot \bm v  \right) - \nu \: \nabla \times \nabla \times \bm v  } 
\label{NSrot3}
\end{eqnarray}

In fact, Stokes' law $3 \: \lambda + 2 \: \mu = 0$ is wrong for fluids, including for monoatomic gas \cite{Gad95b, Raj13}. Dynamic viscosity has real meaning for a viscous fluid, but bulk viscosity or volume viscosity, given by the superposition of both Lam{\'e} coefficients, $\eta = \lambda + 2/3 \: \mu$, has very different values in the literature \cite{Lan59, Cha20, Duk09}, which are usually an order of magnitude greater than $\mu$.

This problem is due to confusion between the effects of viscosity attributed by some authors to viscous dissipation during compression, for example at the level of a shock wave, and the propagation effects of compression waves defined by longitudinal celerity; for the effects of propagation, celerity is written $c_l = 1 / \rho \: \chi$ where $\chi$ is the isothermal or isentropic compressibility coefficient according to the considered evolution. For water, the values are $c_l = 1500 \: m \: s^{-1}$, $\chi = 0.444 \: 10^{- 9} \: Pa^{-1}$ and $\nu = 10^{-6} \: m^2 \: s^{-1}$. By identifying the two definitions of longitudinal celerity, $c_l = \sqrt{(\lambda + 2 \: \mu) / \rho} = 1 / \rho \: \chi$ we find a relation $\lambda / \mu \approx 10^{13}$. As both effects are carried by the same differential operators, $\nabla( \nabla \cdot \bm v)$, viscous dissipation due to compression is completely negligible, at least for large time constants.

The controversies of the last century on this subject \cite{Ros54} and more recent work have not enabled the value of $\lambda$ for fluids to be fixed definitively.  Unlike the values of compression modulus $\lambda$ for solids, which has a certain reality, the values of the first Lam{\'e} coefficient of fluids have no influence on compressive stress $\nabla \left(\left (\lambda + 2 \: \mu \right) \: \nabla \cdot \bm v \right)$. This term is overflowed by an external constraint and this is why the law of conservation of mass $\nabla \cdot \bm v = 0$ is necessarily added here to the equation of motion for an incompressible motion in Navier-Stokes equations. Consequently, the term $\eta \: \nabla \cdot \bm v$ is removed from the equation (\ref {NSrot3}), as this term cannot co-exist with the outer constraint. In discrete mechanics, the importance of each term in the equation (\ref{discrete}) changes according to the time lapse $dt$ with which the phenomenon is observed, and the formulation is adapted to capture physics at all time constants.

For viscosities that vary in space or time for various reasons, it is necessary to consider their spatial locations or what amounts to the same differential operators to which they are appended. The differentiation of physical properties does not mean anything in itself, as it is first necessary to associate an operator with it. For kinematic viscosity $\nu = \mu / \rho$ it is the operator $\nabla \times \bm v$; indeed, it is not so much the rotational velocity that is important but the constraint of rotation-shear $\nu \: \nabla \times \bm v$. The curl of this quantity then represents the acceleration opposed to any change in velocity by the viscous actions.
\begin{eqnarray}
 \left\{  
\begin{array}{llllll}
\displaystyle{  \frac{\partial \bm v}{\partial t} + \frac{1}{2} \: \nabla \left( |\bm v |^2 \right) - \bm v \times \nabla \times \bm v =   - \nabla \phi - \nabla \times \left( \nu \: \nabla \times \bm v \right)  } \\  \\
\displaystyle{ \nabla \cdot \bm v  =  0  }
\end{array} 
\right.
\label{NSrot4}
\end{eqnarray}

The term $\phi = p / \rho$ represents the actions of the forces of pressure per unit of mass; $\phi$ is indeed an energy per unit of mass and its gradient makes it an acceleration. In fact, density is not explicitly represented in the equation of motion. The constitutive equations are not directly present but they intervene through the relation between the pressure and the potential $\phi$. 

\vspace{1.mm}

\subsection{Inertia terms} 

Let us now consider the term of inertia, which, in continuum mechanics, is denoted indifferently $\bm v \cdot \nabla \bm v$ or $ \nabla \cdot \left (\bm v \otimes \bm v \right) - \bm v \: \nabla \cdot \bm v$ or in the form of Lamb (\ref{NSrot4}). They are strictly equivalent and do not affect the result. However, the Lamb vector $\mathcal L = \nabla \times \bm v \times \bm v = - \bm v \times \nabla \times v$ is not a curl, which inhibits any possibility of rendering the Navier-Stokes equation compatible with a Helmholtz-Hodge decomposition \cite{Cal20c}. The essential differences are in fact revealed only on the derivatives of the equations or of certain terms like that of inertia. The divergence and the curl of the material derivative highlights these differences.

\vspace{1.mm}
\paragraph{ Inertia divergence : }

the components of the inertia term are sought within an orthonormal frame of reference $(x, y, z)$. Certain quantities such as velocity, acceleration or the components of the Navier-Stokes equation themselves are defined on the components of this unit vector frame of reference $(\mathbf e_x, \mathbf e_y, \mathbf e_z)$. This is not the case of the curl, for example, which is expressed by space plane in the form $\nabla \times \bm v = \bm \eta : \nabla \bm v$, where $\bm \eta$ is the tensor orientation. Vector $- \bm v \times \nabla \times \bm v$ expresses the negative acceleration exerted on any body whose velocity tends to increase. But we can distinguish two distinct actions of inertia, the direct action carried by the term $\nabla \left (| \bm v |^2/2 \right)$ and an induced action carried by $\mathcal L = - \bm v \times \nabla \times \bm v$. If the direct action is a gradient, the induced action is not a curl action in the sense of the continuous medium. The divergence of the Lamb vector consists of two terms, $\nabla \cdot \mathcal L = \bm v \cdot \nabla \times \bm \omega - \bm \omega \cdot \bm \omega$ where $\bm \omega = \nabla \times \bm v$, the flexion product and the negative enstrophy. An equivalent form of this term leads to the divergence of the material derivative \cite{Cal20a}: 
\begin{eqnarray}
\displaystyle{  \nabla \cdot \left(  \frac{d \bm v}{dt} \right) =    \nabla \cdot \left( \frac{\partial \bm v }{\partial t} \right) +  \nabla^2 \phi_i   + \bm v \cdot \nabla \left( \nabla \cdot \bm v \right) +   \left( \nabla \cdot \bm v \right)^2 - 2 \:  I_2  }
\label{divmat}
\end{eqnarray}
where $I_2$ is the second invariant of tensor $ \nabla \bm v $, which of course also appears for the other classical forms of the inertial term. The last three terms are derived from the divergence of the Lamb vector; even for a velocity with zero divergence these terms are not eliminated, which inhibits the possibility of a Helmholtz-Hodge decomposition of inertia.

In discrete mechanics, with $\nabla \cdot \nabla^d \times (\phi_i \: \mathbf n) = 0$, for a compressible flow or not, we would have: 
\begin{eqnarray}
\displaystyle{  \nabla \cdot \left(  \frac{d \bm v}{dt} \right) =    \nabla \cdot \left( \frac{\partial \bm v }{\partial t} \right) +  \nabla^2  \phi_i    }
\label{divmat2}
\end{eqnarray}

The last term can be associated with the gradient of the scalar potential of the acceleration to form, on the second member of the equation of motion, the Bernoulli potential $\phi_B = \phi + \phi_i$. It is no longer necessary to explain $\nabla \phi_i$, only the definition of the potential differs. From a practical point of view, the simulations carried out with the Bernoulli potential prove to be particularly effective.

\vspace{1.mm}
\paragraph{ Inertia curl :}

the application of the curl operator to the inertial terms of the Navier-Stokes equation leads to the expression:
\begin{eqnarray}
\displaystyle{  \nabla \times \left(  \frac{d \bm v}{dt} \right) =    \nabla \times \left( \frac{\partial \bm v }{\partial t} \right)   + \bm v \cdot \nabla \left( \nabla \times \bm v \right) - \nabla \times \bm v \cdot \nabla \bm v  }
\label{rotmat}
\end{eqnarray}
or, by putting $\bm \omega = \nabla \times \bm v$:
\begin{eqnarray}
\displaystyle{  \nabla \times \left(  \frac{d \bm v}{dt} \right) =  \frac{\partial \bm \omega }{\partial t}   + \bm v \cdot \nabla \bm \omega - \bm \omega \cdot \nabla \bm v  }
\label{rotmat2}
\end{eqnarray}
where $\bm v \cdot \nabla \bm \omega$ represents the advection of the vortex by the velocity field. The term $\bm \omega \cdot \nabla \bm v$ is zero in two dimensions of space because the two components are orthogonal, namely: 
\begin{eqnarray}
\displaystyle{  \nabla \times \left(  \frac{d \bm v}{dt} \right) =  \frac{\partial \bm \omega }{\partial t}   + \bm v \cdot \nabla \bm \omega   }
\label{rotmat3}
\end{eqnarray}

The fact that the result is not the same in two and three dimensions of space leads to the question of the legitimacy of this last term, which owes its existence only to the fact that $\bm v$ is a vector of space. If this vector were projected in each space plane of the trihedron orthogonal to each direction $(x, y, z)$, operator $ \nabla \times \bm v$ would be the curl of this projected vector but also that the component on $\mathbf n$ of vector $\bm v$ of $\mathcal R^3$. For each of the planes, the term $\bm \omega \cdot \nabla \bm v$ would be zero and the expression (\ref{rotmat3}) would be the same whatever the dimension of the space.

In discrete mechanics the primal curl of the material derivative becomes: 
\begin{eqnarray}
\displaystyle{  \nabla \times \left(  \frac{d \bm v}{dt} \right) =    \nabla \times \left( \frac{\partial \bm v }{\partial t} \right)  - \nabla \times \nabla^d \times \left( \phi_i \: \mathbf n \right)  }
\label{rotmat4}
\end{eqnarray}

where, according to the rules for the composition of operators (\ref{calculprop}), the last term has a precise meaning because quantity $\left(\bm \psi = \phi_i \: \mathbf n \right)$ is a pseudo-vector.

\vspace{1.mm}
\paragraph{A new formulation of the Navier-Stokes equations : }

\begin{figure}[!ht]
\begin{center}
\includegraphics[width=7.cm]{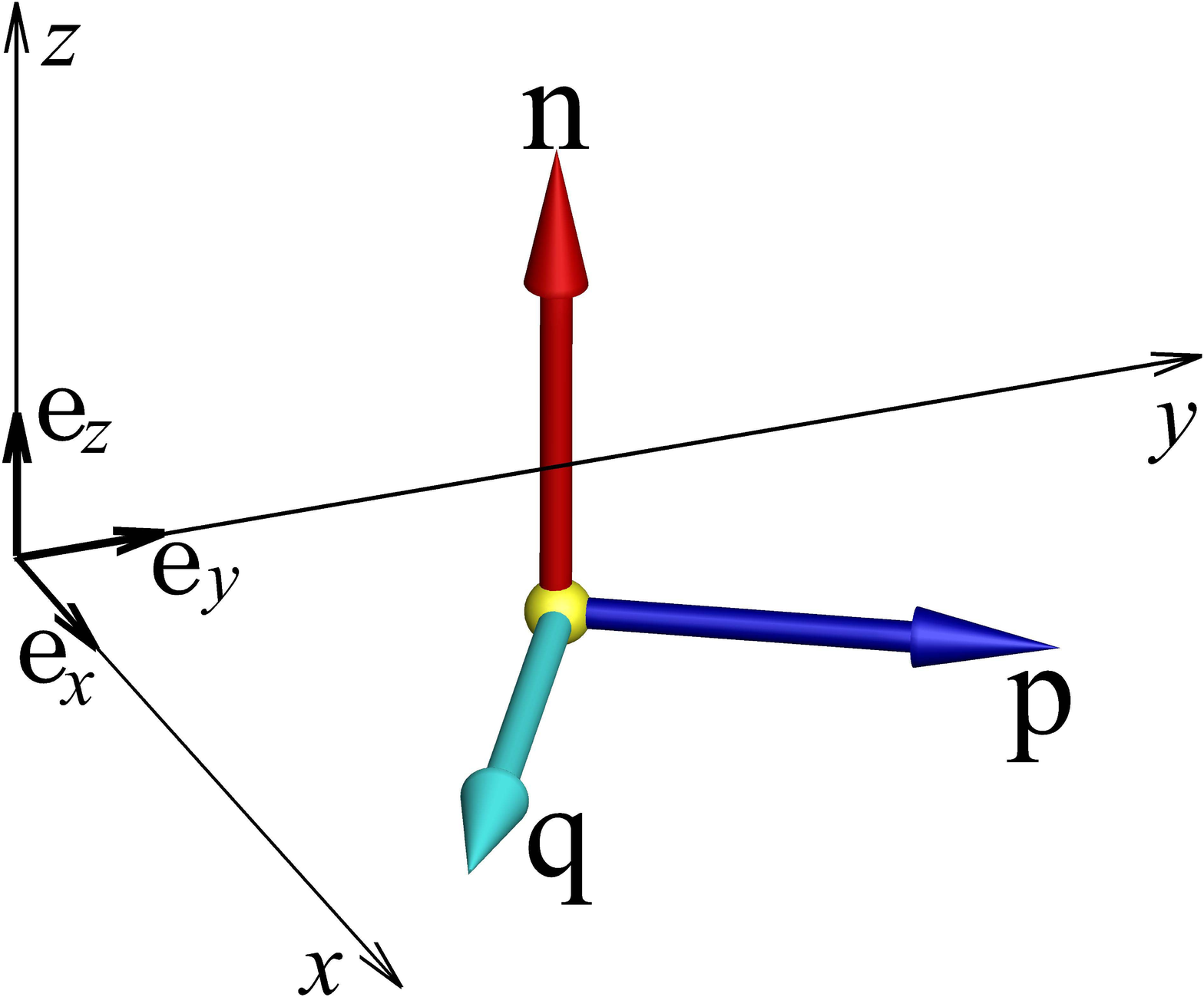}
\caption{ Representation of the vectors of the space plane $(x, y)$, $ \mathbf p = \nabla \left (| \bm v | ^ 2/2 \right) $ and $\mathbf q = \nabla \left (| \bm v |^2/2 \right) \times \mathbf n$ in an orthonormal direct coordinate system where $\mathbf n$ is the direct unit normal orthogonal to this plane. }
\label{inertia}
\end{center}
\end{figure}

In order to establish the expression of the inertia term of the Navier-Stokes equation within the framework of the continuous medium hypothesis, it is necessary to review its physical meaning because it is impossible to directly transform one form of inertia as a sum of a gradient and a curl one. To do so, we must consider the inertial potential $\phi_i = | \bm v |^2/2$, which is defined anywhere in space. The term $\nabla \phi_i$ is the acceleration of the material medium in the plane $(x, y)$ represented in figure (\ref{inertia}) by vector $ \mathbf p = \nabla \phi_i$.

For reasons of symmetry, there is also a vector $\mathbf q$ of the plane $(x, y)$ which is its orthogonal component, $\mathbf q = \nabla \phi_i \times \mathbf n $, where the normal vector is written $\mathbf n = \mathbf e_x \wedge \mathbf e_y$. It is thus the sum $\nabla \phi_i - \nabla \phi_i \times \mathbf n$ which represents the total inertia. This form is consistent with that resulting from discrete mechanics where the inertia takes the form $\nabla \phi_i - \nabla \times (\phi_i \: \mathbf n)$. The formula of vector calculus (\ref{calculus}) transforms this last term in the form:
\begin{eqnarray}
\displaystyle{  \frac{1}{2} \: \nabla \times \left( |\bm v |^2 \: \mathbf n \right) =  
\frac{1}{2} \: \nabla \left( |\bm v |^2 \right) \times \mathbf n + \frac{1}{2} \: |\bm v |^2 \nabla \times \mathbf n  }
\label{calculus}
\end{eqnarray}

But here, unlike in discrete mechanics, the unit vector oriented normal to the space plane $(x, y)$ is constant and $\nabla \times \mathbf n = 0$. The first term of the second member of equality (\ref{calculus})  must be calculated by plane, then projected on its coordinates. Thus the inertia terms are written $\nabla \phi_i - \nabla \phi_i \times \mathbf n$, in two orthogonal terms, while respecting the rules defined in continuous medium.
The new form of the Navier-Stokes equations is then written:
\begin{eqnarray}
 \left\{  
\begin{array}{llllll}
\displaystyle{  \frac{\partial \bm v}{\partial t} + \frac{1}{2} \: \nabla \left( |\bm v |^2 \right) - \frac{1}{2} \: \nabla \left( |\bm v |^2 \right) \times \mathbf n =   - \nabla \phi - \nabla \times \left( \nu \: \nabla \times \bm v \right)  } \\  \\
\displaystyle{ \nabla \cdot \bm v  =  0  }
\end{array} 
\right.
\label{NSnew}
\end{eqnarray}
where the notation of the second inertia term $\nabla \left (| \bm v |^2/2 \right) \times \mathbf n$ corresponds to the gradient orthogonal to $\nabla \left (| \bm v |^2/2 \right)$ in the plane orthogonal to $\mathbf n$. This continuous medium type notation must then be interpreted by component within the framework of the Galilean frame of reference.

The inertial term $\nabla \phi_i$ is indeed zero curl, but this new form of the Navier-Stokes equation (\ref{NSnew}) also leads to $ \nabla \cdot \left (\nabla \phi_i \times \mathbf n \right) = $ 0. Indeed, vectors $\mathbf p$ and $\mathbf q$ are orthogonal $\mathbf p \cdot \mathbf q = 0$; vector $\mathbf p$ represents the trajectories of the sources and sinks of the flow, while vector $\mathbf q$ defines the closed streamlines corresponding to the inertial vector potential.

The two terms of inertia are generally different from zero but their sum can possibly be zero. This is the case with Couette and Poiseuille flows, where each of the two contributions is different from zero but where the global inertia is zero; the current lines corresponding to these flows are parallel rectilinear straight lines. In fact, the quantity:
\begin{eqnarray}
\displaystyle{  \kappa_i =  \frac{ \nabla \phi_i  - \nabla \phi_i \times \mathbf n }{ | \phi_i |} }
\label{curvature}
\end{eqnarray}
defines the local curvature of the inertial potential. This quantity reflects the changes in direction of the fluid, for example in the presence of an obstacle, and becomes potentially infinite at breakpoints.

\vspace{1.mm}
A difference between the Navier-Stokes equation (\ref{NSnew}) and the discrete equation of motion (\ref{discrete}) is in the writing of the second inertia term, with respectively $- \nabla \phi_i \times \mathbf n $ and $\nabla \times \left (\phi_i \: \mathbf n \right)$, only the definition of the unit vector $ \mathbf n $ differs; however, they are strictly equivalent. The real modification concerns the Lamb vector $ \mathcal L = - \bm v \times \nabla \times \bm v $, which is not divergence-free. The second invariant $I_2$ of the tensor $\nabla \bm v$ is a second order term of the spatial increment with which it is evaluated. It appears as a compatibility condition to be satisfied when solving the standard Navier-Stokes equations \cite{Cal20c}. The form (\ref{NSnew}) removes this condition and allows the equation of motion to be a true Helmholtz-Hodge decomposition. If, from a practical point of view, the consequences are minor, the formal differences are important, in particular in the application of the operator divergence to the equation of motion for the projection methods \cite{Gue06} or the mathematical properties of regularity of three-dimensional space solutions.

\subsection{Inertia in rotating uniform flow } 

Consider a uniform rotational flow of velocity $\Omega \cdot \mathbf ez = \omega_0$, constant in a cylindrical coordinate system $(r, \theta, z)$. This flow is stationary incompressible and its curl is constant, equal to $\nabla \times \mathbf V = 2 \: \omega_0 \: \mathbf e_z$ and, as it is constant, its curl is zero $\nabla \times (\nu \: \nabla \times \mathbf V) = 0$; the viscous effects are thus equal to zero. In addition, $ \nabla \cdot \bm v = 0 $ and the pressure (or $\phi$) will remain equal to zero, and the movement is to be considered only from the kinematic point of view.

The only component of the non-identically zero Navier-Stokes equation in its classical version is that on $r$, it is written $\rho \: V _ {\theta}^2 / r = 0$ with $V_{\theta} = \omega_0 \: r$ or $\rho \: \omega_0^2 \: r = 0$; the mechanical equilibrium is therefore not guaranteed. This term corresponds to the non-zero inertia of the Navier-Stokes equation for $\omega_0$ other than zero. To obtain the mechanical equilibrium corresponding to this flow, it is necessary to add a fictitious centripetal force to the equation; inertia is not intrinsically zero.

Let us now take the discrete equation of motion or the new form of the Navier-Stokes equation where the inertia is written $\nabla (| \bm v | ^ 2/2) - \nabla (| \bm v |^2/2) \times \mathbf n$, with here $\mathbf n = \mathbf e_z$ and $\nabla (| \bm v |^2/2) = \omega_0^2 \: r$. In this case the inertia is strictly null locally by compensation of the two terms.

\vspace{1.mm}
To conclude, the Navier-Stokes equations in their classical forms (\ref{navier}) present a paradox directly inherited from the concept of continuous medium, where the mechanical equilibrium must imperatively be verified on each of the components of a global frame of reference. The new form (\ref{NSnew}) expresses the inertial term in coherence with that of the discrete mechanics, while remaining within the assumption of continuous medium.

\section{Analysis of steady flow past a circular cylinder at $Re = 30$ } 

The equation of discrete mechanics (\ref{discrete}) is retained to simulate an inertial stationary incompressible flow used as a standard test case in computational fluid mechanics. The objective is to find the results of the literature in this example, but above all to highlight the role of the inertial potential.

The example presented is a simple illustration of the direct use of the discrete formulation. It involves calculating the solution $(\phi, \bm \psi, \bm v)$ corresponding to a very classical test case of fluid mechanics, a steady flow past a circular cylinder for a Reynolds number $Re = V_0 \: D / \nu$, where $V_0$ is the velocity at infinity, $ D $ the diameter of the cylinder and $\nu$ the kinematic viscosity. For Reynolds numbers less than about $ Re = 42 $, the flow is steady and has two dimensions of space. The characteristics of the flow, recirculation length, drag and separation angle, have been precisely determined by many authors, including \cite{Sen09} and \cite{Lam13} in particular for a Reynolds number $Re = 40$ close to the unsteady transition. For a range of Reynolds numbers $0.6 <Re \leq 42$, the flow has a recirculation of dimensionless length $ L = 2 \: \bm l / D $ where $ \bm l $ is the actual length of the recirculation.

The objective is not to find more precise solutions than those already present in the literature but to present the specificities of the discrete model through this example for a value of the Reynolds number $ Re = 30 $. For this Reynolds number the value of the recirculation varies from $ L = 1.3 $ to $ L = 3.16 $ depending on the authors, but the most recent studies give a recirculation close to $L = 3.16$. In fact, a certain number of parameters influence this characteristic of the flow: the lateral and longitudinal extensions of the domain of resolution, the number of degrees of freedom of the mesh, the quality of the mesh, etc. We must probably add more profound reasons linked to the representativeness of the Navier-Stokes equation when the inertial effects tend towards zero; the Stokes paradox for this problem is indeed found when one deviates very far from the cylinder where the inertial effects disappear.
\begin{figure}[!ht]
\begin{center}
\includegraphics[width=8.cm]{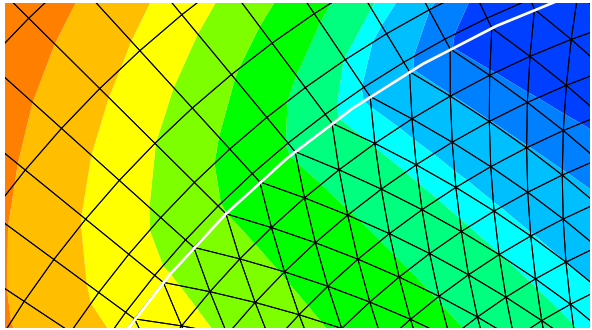}
\caption{ Detail of mesh composed by $n =128 \times 64$ quadrangles in fluids and $n = 8192$ triangles in solid }
\label{mesh}
\end{center}
\end{figure}

As the objective is the analysis of the flow in terms of potentials, let us consider this test case and its stationary solution obtained from the discrete equation (\ref{discrete}). The solution will not only be sought in the fluid domain but also in the solid cylinder, where the velocity is zero by definition but not the pressure or shear stresses. Several structured and unstructured meshes have been produced, but a detail of the one selected is presented in figure (\ref{mesh}). This non-regular conformal hybrid mesh is composed of a structured mesh part conforming to the cylinder and an unstructured mesh based on rather uniform triangles for the solid cylinder.

The two meshes can be united in a single hybrid mesh to find the solution in the two domains by ensuring a zero velocity $\bm v = 0$ on circle $\Gamma_c$ and by fixing a value of mobility $\kappa$ large enough to ensure a velocity of order of magnitude of machine precision \cite{Kha00}. Another solution is to calculate the fluid solution and solve the equations (\ref{darcy}) to determine the potentials in the solid domain; in fact the solution in the fluid does not depend on that of the solid.
\begin{figure}[!ht]
\begin{center}
\includegraphics[width=10.cm]{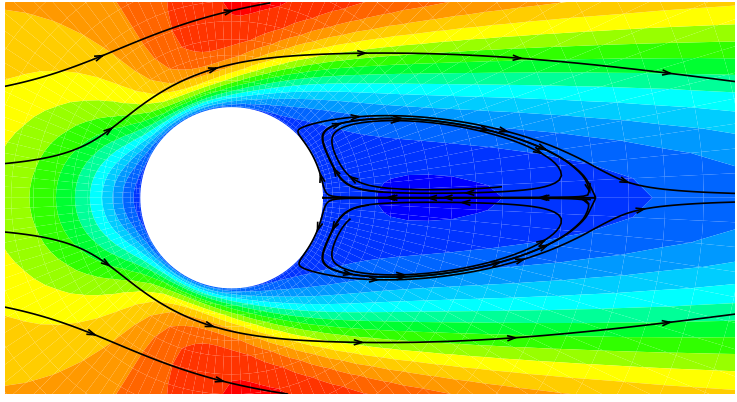}
\caption{ Streamwise velocity and trajectories in the recirculation past cylinder. }
\label{cylinder-vtx}
\end{center}
\end{figure}

Figure (\ref{cylinder-vtx}) shows the solution obtained in the immediate vicinity of the cylinder in the case where only the fluid is taken into account in the simulation.
The recirculation length is equal to $L = 2.98$, a value close to the current state of the art. As expected, the solution is symmetrical with respect to a horizontal axis and the regime is perfectly steady.
The two symmetric, counter-rotating recirculation zones remain stable, attached to the cylinder at separation angle $\theta_s$ close to $50^o$. For the Reynolds number considered, the near-wall mesh is sufficiently fine to capture the boundary layer and to correctly represent the shear stresses in this zone.

\begin{figure}[!ht]
\begin{center}
\includegraphics[width=5.cm]{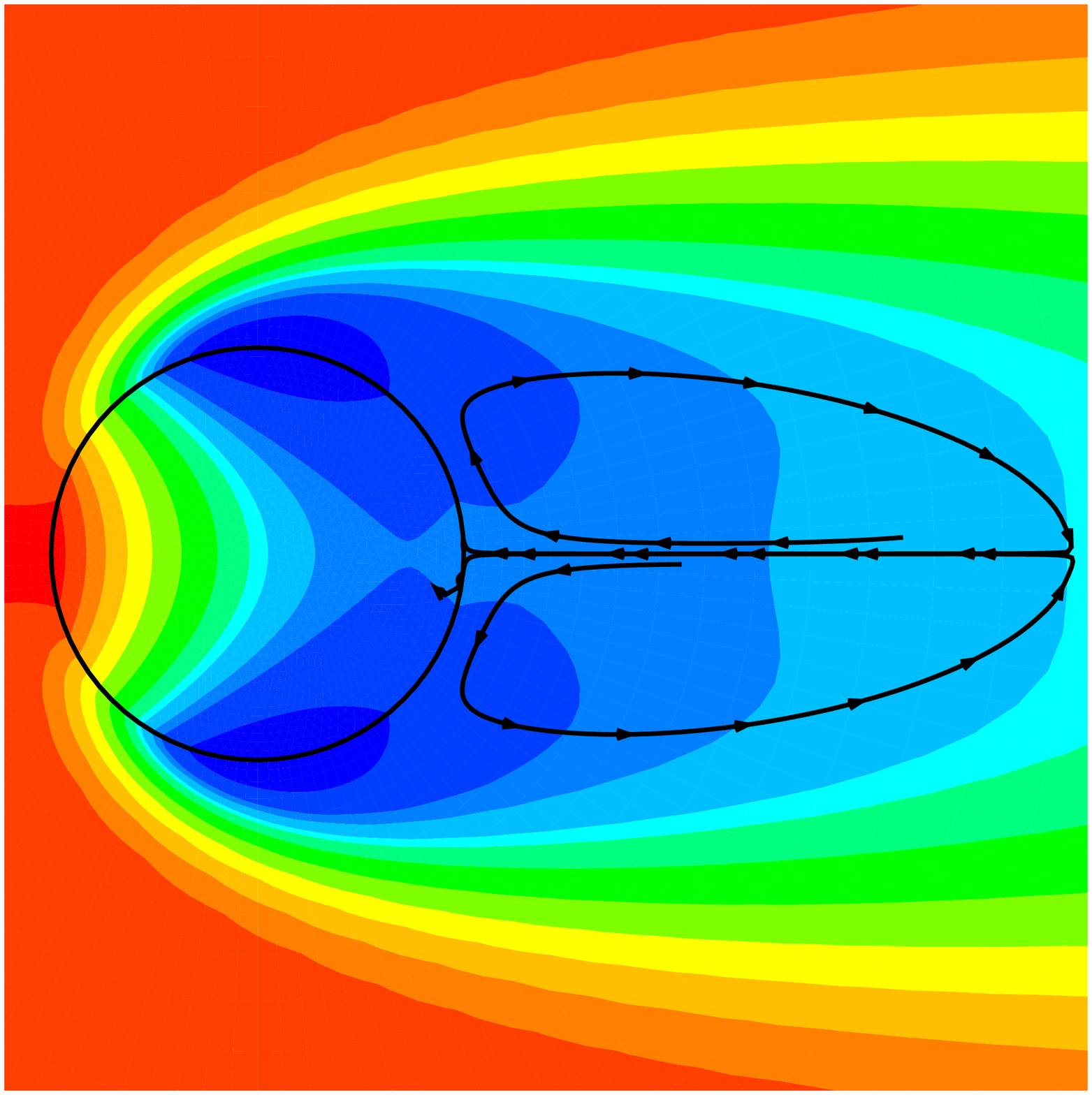}
\includegraphics[width=5.cm]{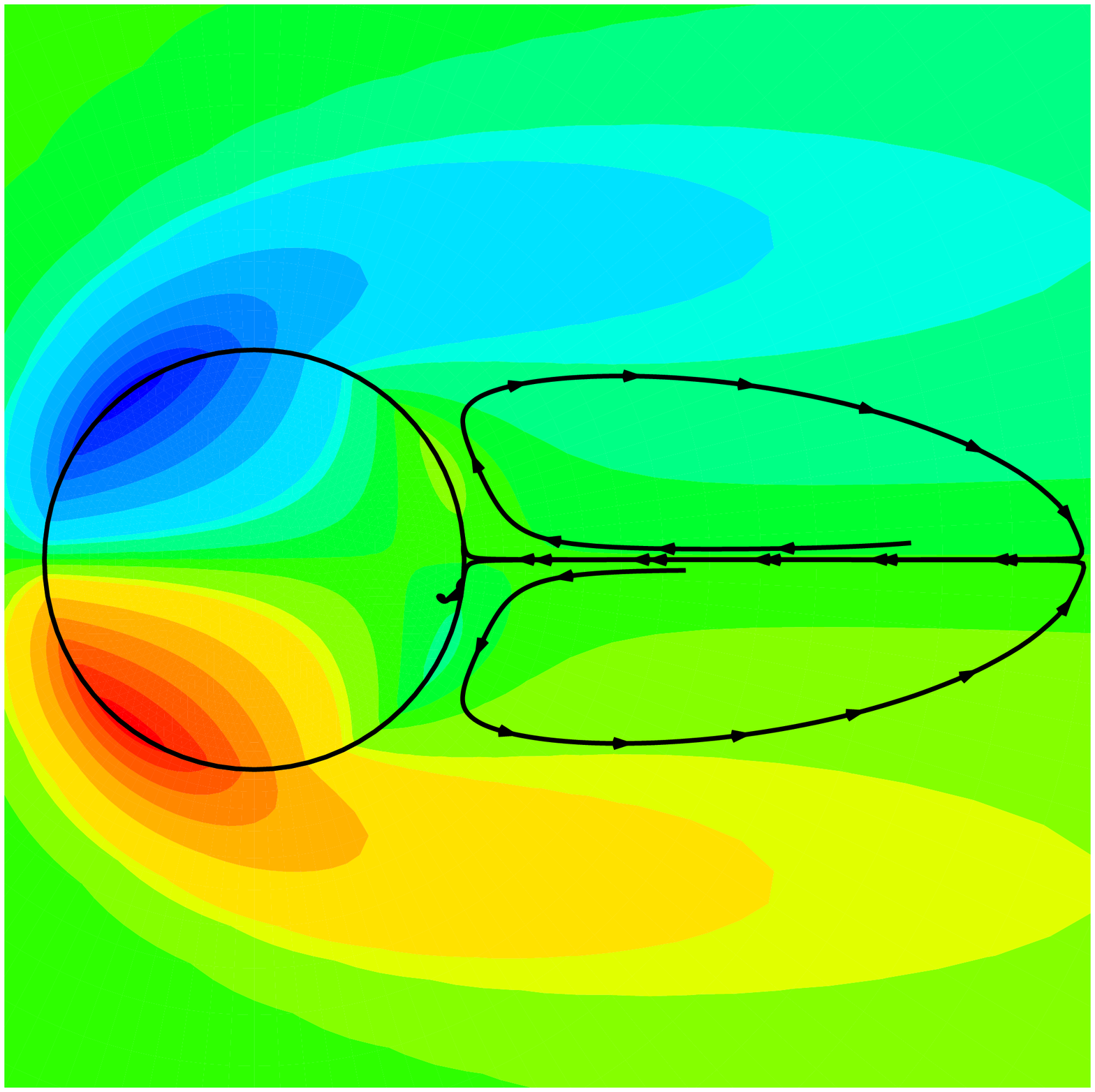}
\includegraphics[width=5.cm]{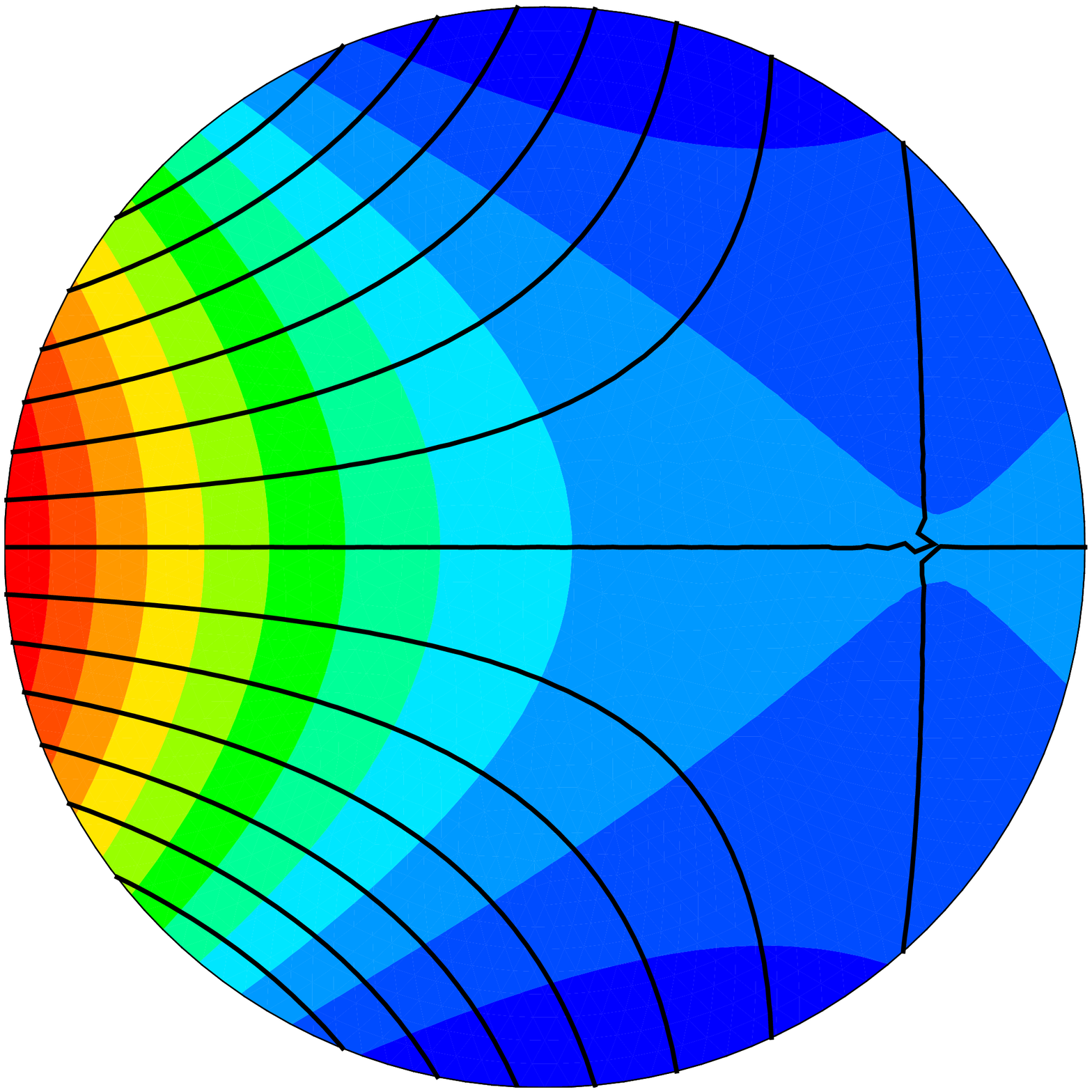}
\caption{ Scalar and  vector potentials $\phi^o$ and $\bm \psi^o$ in fluid and solid. To right, a detail of scalar potential $\phi^o_s$ field and vector potential $\bm \psi^o_s$ (lines) in solid cylinder.}
\label{cylinder-phipsi}
\end{center}
\end{figure}

Solving equations $\nabla^2 \phi_s = 0$ and $\nabla^2 \bm \psi_s = 0$ is performed separately in the solid domain $\Omega_s$ from the solution of potentials $\phi_{\Gamma_s} $ and $\bm \psi_{\Gamma_s} $ on the trace of cylinder $\Gamma_s$. Figure (\ref{cylinder-phipsi}) shows the complete solution in both domains for the two potentials.
When the resolution in the two domains is coupled, the velocity in the solid is of order of magnitude of $ 1 / \kappa $ and it is strictly zero when the potentials in the solid are calculated separately. The velocity field in the solid area can be calculated {\it a posteriori} from:
\begin{eqnarray}
\displaystyle{ \bm v = - \frac{1}{\kappa} \: \nabla \phi_s = \frac{1}{\kappa} \: \nabla^d \times \bm  \psi_s }
\label{vitdarcy}
\end{eqnarray}

Both fields equal, $- \nabla \phi_s = \nabla^d \times \bm \psi_s$, but they are also orthogonal, $\nabla \phi_s \cdot \nabla^d \times \bm \psi_s = 0$ as shown in figure (\ref{cylinder-phipsi}). The two fields are harmonic, $\nabla^2 \phi_s = 0$ and $\nabla^2 \bm \psi_s = 0$.

The potential fields $ \phi_s$ and $\bm \psi_s$ make it possible to calculate the compression and shear stresses on the trace of the cylinder $\Gamma_s$ from the fluid field, but it is also possible to calculate these stresses at the interior of the solid as surface integrals:
\begin{eqnarray}
\displaystyle{ \bm f_c = - \int_{\Omega_s} \nabla \phi_s \: ds \:\:\:\:\:\:\:\:\:\:\:\: \bm f_s = \int_{\Omega_s}   \nabla^d \times \bm  \psi_s \: ds}
\label{forces}
\end{eqnarray}

\begin{figure}[!ht]
\begin{center}
\includegraphics[width=10.cm]{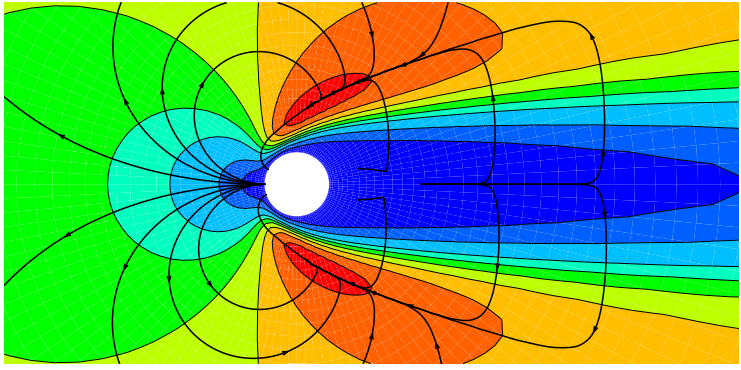}
\caption{ Field of $\mathbf q = \nabla^d \times \left( \phi_i \: \mathbf n \right)$ and streamlines of field $\mathbf p = \nabla \phi_i$  with $\phi_i =| \bm v |^2 / 2$.  }
\label{analyseR2V2}
\end{center}
\end{figure}

The inertia terms $\nabla \phi_i$ and $\nabla^d \times \left( \phi_i \: \mathbf n \right)$ can be evaluated from the inertial potential $\phi_i$; these fields are shown in figure (\ref{analyseR2V2}).

The maxima of potential $\phi_i$ are located symmetrically on each side of the solid cylinder and slightly shifted downstream of the flow; they correspond to the wells of $ \nabla \phi_i $ while the main source is located at the breakpoint of the flow in $\theta = 180^o$. These zones with high inertial potential correspond to those where the product of the curvature by potential $\kappa_i \: | \phi_i |$ are maximum. As expected, the current lines of the potential of $\mathbf q = \nabla \times \bm \psi_i $ are orthogonal in all points with those of $\mathbf p = \nabla \phi_i$.

This example of inertial incompressible stationary flow shows that the classical results of the literature can be replicated by the discrete formulation (\ref{discrete}). The numerous examples of flows simulated with this equation, moreover, corroborate the fact that this equation of motion makes it possible to find the results acquired with the Navier-Stokes equation in its classical form. The new formulation of the Navier-Stokes equations (\ref{NSnew}) has exactly the same properties as the equation resulting from discrete mechanics. However, profound fundamental differences persist between the proposed formulation and the classical form of the Navier-Stokes equations (\ref{navier}).   

The performances of the methodologies resulting from the discrete formulation used or from the new form of the Navier-Stokes equations are very comparable from a technical point of view, in terms of calculation time, memory space, etc. The discrete mechanics methodology shows the same properties as the Augmented Lagrangian method \cite{For82}, with slightly higher computation times than the projection methods and very great robustness due to integration of the incompressibility constraint in the equation of motion. The system of equations (\ref{discrete}) is directly transposable to an algebraic version, since all the operators express the velocity component on a segment. The formulation merges with the numerical method. For the new form of the Navier-Stokes equations, we note the return to the concept of continuous medium and its writing in two contributions, one divergence-free and the other curl-free. The application of the conservation of mass as an adjoint equation allows a free choice of the coupling method between the scalar potential and the velocity.

\section{Conclusions} 

The formulation of the Navier-Stokes equations proposed here presents multiple advantages, namely (i) the absence of tensors of order equal to or greater than two, replaced by simple differential operators, (ii) the absence of derivation of physical properties when these are spatially variable, (iii) the absence of density in the equation of motion, replaced by the scalar potential $ \phi $, (iv) the definition, as energies per unit mass, scalar and vector potentials of the acceleration resulting from the formal Helmholtz-Hodge decomposition of the equation, (v) the new formulation of the inertial term in a sum of a curl-free term and another divergence-free term, and (vi) perfect continuity of scalar and vector potentials of acceleration defined in fluids, solids and porous media. The absence of density within the discrete equation expressed in terms of accelerations would extend its meaning to the propagation of electromagnetic waves in a vacuum.

The exact or approximate solutions of the Navier-Stokes equations in its new form (\ref{NSnew}) and of the discrete equation of motion (\ref{discrete}) are now exactly the same. Certain characteristics inherited from the discrete equation and transposed to the Navier-Stokes equations give it particular properties that can be used directly for the simulation of flows or for its mathematical analysis within the framework of the concept of continuous medium.

\vspace{4.mm}

{\bf Author Contribution}
\vspace{2.mm}

Author has performed Physical modeling, Conceptualization, Methodology, Research code,  Validation, Writing-Original draft preparation, Reviewing and Editing.

The paper has been checked by a proofreader of English origin.

\vspace{2.mm}
{\bf Data Availability}
\vspace{2.mm}

The data that support the findings of this study are available within the article.

\vspace{2.mm}

{\bf Declaration of competing interest}
\vspace{2.mm}

There are no conflict of interest in this work.


\end{document}